\documentstyle[emulateapj,psfig,epsf]{article}
\def\kms{km$\,{\rm s}^{-1}$}
\def\etal{\it et~al.\rm}
\def\sb{ergs cm$^{-2}$ s$^{-1}$ arcsec$^{-2}$}
\def\fl{ergs cm$^{-2}$ s$^{-1}$}
\begin{document}
\title{The Optical Spectrum of the SN 1006 Supernova Remnant Revisited}
\submitted{Submitted January 31, 2002}

\author{Parviz Ghavamian\altaffilmark{1}, P. Frank Winkler\altaffilmark{2,5},
John C. Raymond\altaffilmark{3} and Knox S. Long\altaffilmark{4,5} }

\begin{abstract}

We present the deepest optical spectrum acquired to date of Balmer-dominated shocks in the NW
rim of SN 1006.  We detect the broad and narrow components of H$\alpha$, H$\beta$
and H$\gamma$ and report the first detection of the He~I $\lambda$ 6678 emission line
in this supernova remnant.  We may have detected, at the 1.5$\sigma$ level, faint He~II $\lambda$4686 
emission.  We measure a full width half maximum of 2290\,$\pm$\,80 \kms\,
in the broad component H$\alpha$ line, with broad-to-narrow flux ratios of 0.84$^{\,+0.03}_{-0.01}$
and 0.93$^{+0.18}_{-0.16}$ in H$\alpha$ and H$\beta$, respectively.  To match these observations,
our nonradiative shock models
require a low degree of electron-proton equilibration at the shock front, $T_{e}/T_{p}\,\leq\,$
0.07, and a shock speed of 2890$\pm$100 \kms.  These results agree well with 
an earlier analysis of ultraviolet lines from SN 1006.  The He~I/H$\alpha$ and He~I/He~II flux
ratios also indicate low equilibration.  Furthermore, our models match the observations for mostly ionized
($\sim\,$90\%) preshock H and mostly neutral ($\gtrsim\,$70\%) preshock He, respectively.
We conclude that the high H ionization fraction cannot be explained
by either photoionization from the reverse shock or relic ionization from EUV photons released in the 1006 A.D.
supernova.  The most plausible explanation appears to be photoionization from the Galactic Lyman
continuum.  

\keywords{ ISM: supernova remnants: individual (SN 1006)--ISM: kinematics and dynamics, shock waves}
\end{abstract}
\altaffiltext{1}{Department of Physics and Astronomy, Rutgers University, 136 Frelinghuysen
Rd., Piscataway, NJ 08854-8019; parviz@physics.rutgers.edu}

\altaffiltext{2}{Department of Physics, Middlebury College, Middlebury, VT 05753; 
winkler@middlebury.edu}

\altaffiltext{3}{Harvard-Smithsonian Center for Astrophysics, 60 Garden St., Cambridge, MA
02138; jraymond@cfa.harvard.edu}

\altaffiltext{4}{Space Telescope Science Institute, 3700 San Martin Dr., Baltimore, MD 21218,
long@stsci.edu}

\altaffiltext{5}{Visiting Astronomer, Cerro Tololo Inter-American Observatory, National Optical Astronomy
Observatories.  CTIO is operated by AURA, Inc.\ under contract to the National Science
Foundation.}

\section{INTRODUCTION}

Among the known galactic supernova remnants (SNRs), SN 1006 stands out as one of the few with low extinction ($E(B\,-\,V)\,=\,$0.11,
Schweizer \& Middleditch 1980), a well determined age and a reasonably well
known distance ($D\,\sim\,$2 kpc, Kirshner, Winkler \& Chevalier 1987 (hereafter KWC87), Long \etal\,1988, 
Laming \etal\,1996, Winkler \&
Long 1997).  This SNR was probably produced by a Type Ia explosion (Schaefer 1996) and appears as a limb brightened shell 30$^{\arcmin}$
in diameter at X-ray (Willingale \etal\,1996, Winkler \& Long 1997) and radio (Reynolds \& Gilmore 1986) wavelengths.  Using the
Hopkins Ultraviolet Telescope (HUT), Raymond \etal\,(1995) first detected ultraviolet emission lines of He~II $\lambda$1640,
C~IV $\lambda\lambda$1548, 1551, N~V $\lambda\lambda$1239, 1243 and O~VI $\lambda\lambda$1032, 1038 in SN 1006. 
Raymond \etal\,(1995) observed equal velocity widths for the emission lines detected
by HUT. Combining this information with the relative strengths of the He~II, C~IV, N~V and O~VI lines, Laming \etal\,(1996)
concluded that the
lines observed by HUT were due to collisional excitation in 2600 \kms\, nonradiative shocks with
little or no electron-ion temperature equilibration at the shock front.  In this case the high shock speed and low density
($n\,\lesssim\,$ 1 cm$^{-3}$) result in negligible radiative cooling downstream (hence the term `nonradiative shock').
All of the line emission is produced in a narrow ionization zone close to the shock front.

In the optical, SN 1006 appears as a faint network of Balmer-dominated filaments (van den Bergh 1976, Long, Blair \&
van den Bergh 1988, Winkler \& Long 1997). 
These filaments emit an optical spectrum dominated by the Balmer lines of H, produced in the same ionization zone as
the ultraviolet lines.  Balmer line profiles consist of two components: a narrow component produced when
cold ambient neutrals are overrun by the shock and a broad component produced when charge exchange with protons 
generates a hot neutral population behind the shock (Chevalier \& Raymond 1978, Chevalier, Kirshner \& Raymond 1980). 
The broad component width is directly proportional to the postshock proton temperature, while the ratio of broad to narrow flux 
is sensitive to the electron-proton temperature equilibration and the preshock neutral fraction.  These diagnostics
are invaluable tools for measuring shock speeds in supernova remnants and studying the efficiency of collisionless
heating processes in high Mach number, nonradiative shocks.

\section{SPECTROSCOPIC OBSERVATIONS}

We obtained a deep long-slit spectrum of the NW filament in SN~1006 from
the 4-m Blanco telescope at CTIO on 1998 June 24 (UT).  The RC Spectrograph
was used with the 527 line mm$^{-1}$\ grating KPGL3, the Blue Air Schmidt 
camera, and the Loral 3K CCD.  This combination covered the spectral 
range 3760$-$7480 \AA\,at a dispersion of 1.22 \AA\, pixel$^{-1}$.  For our slit width of  2\farcs0, 
the instrumental resolution was approximately 4.5 \AA\,(205 \kms\,at H$\alpha$) along the entire spectrum. 
The plate scale for this configuration was 0\farcs51 pixel$^{-1}$ and the slit length was 5$^{\arcmin}$.
We rotated the slit to PA\,=\,63.9$^{\circ}$ and placed its center at coordinates 15$^{\rm h}$02$^{\rm m}$13\fs 5,
$-$41$^{\circ}$45$^{\arcmin}$22$^{\arcsec}$ (2000), as shown
in Fig.~1.  We chose this position to cut obliquely across the NW
filament in order to capture as much of the bright region as possible 
while giving enough background sky, away from the filament, for a
good sky subtraction.  It is virtually identical to the position
used by KWC87.  We obtained 4 exposures at this location, with a total exposure time
of 8400 s, which we combined into a single two-dimensional spectrum following initial reduction.

The data were reduced using standard IRAF\footnote{IRAF is distributed by the National Optical Astronomy
Observatories, which is operated by the AURA, Inc. under cooperative agreement with the National Science Foundation}
procedures of bias subtraction, flat fielding and illumination correction.  Spectra of a HeNeAr lamp taken at the beginning and end of the series
of object frames were used for wavelength calibration.   We subtracted the night sky contribution from the two-dimensional
spectrum using emission from the two ends of the slit. The H$\alpha$ line profiles (both broad and narrow components)
were nearly constant along the observed filament; therefore, we integrated the emission along the slit
to obtain a single high signal-to-noise spectrum from a 51$^{\arcsec}$ section of the filament (Figs. 2 and 3).  Exposures of
several spectrophotometric standard stars from Hamuy et al. (1994)
were used for flux  calibration.   We estimate an absolute photometric accuracy of
20\% for our quoted emission line fluxes. 

\section{DETECTED EMISSION LINES}

Among the most interesting new features in our optical data are the detection of broad H$\beta$
and He~I $\lambda$ 6678.  To our knowledge, this is the first detection of
the $\lambda$6678 line in a pure nonradiative shock.  
The neutral He atoms passing downstream are unaffected by the electromagnetic turbulence at the shock front.  
Therefore, neutral He atoms remain cool throughout the shock and produce narrow emission lines unresolved in our data.
On the other hand since He$^{+}$ is an ion, it is heated at the shock front by the same collisionless processes
which heat the electrons and ions.  Therefore, we expect the He~II $\lambda$4686 line to be broad
and more difficult to detect than the He~I $\lambda$6678 line.  There is a hint of the
He~II $\lambda$4686 line at a low (1.5$\sigma$) statistical significance.
The surface brightnesses of the detected emission lines from NW SN 1006 appear in Table~1.
We have used $E(B-V) = 0.11$\ (Schweizer \& Middleditch 1980) 
to obtained dereddened line intensities from our measured spectra.

\section{SPECTRAL LINE FITS}

We fit the line profiles using a standard nonlinear $\chi^{2}$ reduction routine.  We assumed
Gaussian profile shapes for all lines and fitted the H$\alpha$, He~I, H$\beta$, H$\gamma$ and He~II lines separately.
Each Balmer line profile was fit with two Gaussians, one each for the narrow and broad components.  The
He lines were fit with single Gaussians.
To reduce the required degrees of freedom, we set the widths of the narrow component Balmer 
lines and He~I line equal to the instrumental resolution during the fitting process.  

Due to the faintness of the $\lambda$4686 emission, we cannot estimate the width of
this line from the fits themselves. Therefore, we set the velocity width of the He~II emission line width equal to that
of the H$\alpha$ line.  This assumption is supported by the HUT analysis of Raymond \etal\,(1995), who found 
that to within the errors, the velocity widths of the He~II $\lambda$1640 and broad Ly $\beta$ lines are equal 
(i.e., electron-ion and ion-ion temperature equilibration is inefficient in SN 1006).
The best fit for the He~II $\lambda$4686 line profile appears in Fig.~4.
We note that due to the large width and faintness
of the He~II line, the flux predicted by the line fit is quite sensitive to the baseline level:
the lower the baseline, the higher the predicted He~II flux.  Both the statistical uncertainty
and the systematic baseline uncertainty contribute to the large error bars on the 
flux.  For the purpose of our subsequent analysis, the most important quantity here is the upper limit
on the He~II line flux.

Our best fit for the H$\alpha$ broad component width is
2290$\pm$80 \kms, agreeing very well with the value of 2310$\pm$210 \kms\,measured by Smith \etal\,(1991).  The
line center shift between the broad and narrow component lines is 29$\pm$18 \kms, less than the uncertainty
in broad component width, indicating a nearly edge-on viewing angle ($\mid 90^{\circ} - \Theta \mid\,\leq\,$2$^{\circ}$). 
The H$\alpha$ broad-to-narrow ratio is 0.84$^{\,+0.03}_{-0.01}$, lying slightly above the extreme upper limit
quoted by Smith \etal\,(1991), 0.73$\pm$0.06.  The H$\beta$ broad-to-narrow ratio is
somewhat larger, 0.93$^{+0.18}_{-0.16}$.  The H$\beta$ broad-to-narrow ratio is larger than that of H$\alpha$
due to the lower conversion efficiency
of Ly $\gamma$ photons into H$\beta$ in the narrow component (c.f. Ghavamian \etal\,2001).  The broad-to-narrow ratio of the 
H$\gamma$ line (0.63$^{+0.53}_{-0.25}$) is poorly constrained due to the faintness of the broad component.  The observed
H$\alpha$ surface brightness of the Balmer filament in SN 1006 is (3.2$\pm$0.7)$\times$10$^{-16}$ \sb\,
(including the 20\% photometric uncertainty), in reasonable agreement with the value quoted by KWC87.

\section{ANALYSIS AND MODELS}

To model the Balmer line profiles, we used the one-dimensional, plane-parallel shock
code described in Ghavamian \etal\,(2001) to compute a grid of numerical models.  Briefly, the
code predicts the density and temperature of electrons, protons and hot neutrals behind a nonradiative
shock.  Broad to narrow flux ratios are computed by a Monte Carlo simulation which follows collisionally excited
Ly $\beta$ and Ly $\gamma$ photons until they are absorbed or escape from the grid.  Collisional excitation
and ionization of H by both electrons and protons is included in the models.
Free parameters of the model are the preshock neutral fraction $f_{H^{0}}$, shock speed, $v_{S}$, and fractional
equilibration immediately behind the shock.  The latter quantity denotes the degree of electron-ion
equilibration by plasma turbulence at the shock front, where $f_{eq}$\,=\,0 represents no (pure Coulomb) equilibration, 
while $f_{eq}$\,=\,1 represents full equilibration.
The electron and proton temperatures are then defined by the relation
\begin{equation}
T_{e}\,=\,\frac{3}{16}\frac{m_{p}\,v_{S}^{2}}{k}\,(\mu\,f_{eq}\,+\,\frac{m_{e}}{m_{p}}\,(1\,-\,f_{eq}))
\label{equile}
\end{equation}
\begin{equation}
T_{p}\,=\,\frac{3}{16}\frac{m_{p}\,v_{S}^{2}}{k}\,(\mu\,f_{eq}\,+\,(1\,-\,f_{eq}))
\label{equilp}
\end{equation}
where $\mu$ is the mean molecular weight (\,=\,0.6 for Galactic abundances, Allen 1973) and $\frac{3}{16}\mu\,m_{p}\,v_{S}^{2}$
is the mean postshock temperature.  For an input $f_{eq}$,
the subsequent evolution of $T_{e}$ and $T_{p}$ by Coulomb collisions is followed as a function of 
position behind the shock.  We have updated the code
to include the calculation of both the He ionization balance and the contribution of 
He ionization to the postshock electron density.  To increase the signal-to-noise of our simulations,
we have also increased the number of excitations
followed by the Monte Carlo simulation from 10,000 used in the earlier paper (Ghavamian \etal\,2001)
to 50,000.

Using the same method outlined in Ghavamian \etal\,(2001),
we computed a grid of models in equilibration and shock velocity, combining each 
$f_{eq}$ with the $v_{S}$ required to match the observed FWHM of the broad component.
We sampled the subset $f_{eq}$\,=\,(0, 0.03, 0.05, 0.1, 0.2, 0.4, 0.6, 0.8, 1), corresponding
to $v_{S}$\,=\,(2865, 2880, 2895, 2920, 2965, 3100, 3235, 3395, 3580) \kms, respectively.  The
H$\alpha$ and H$\beta$ broad-to-narrow ratios predicted by our models are shown in Fig.~5.  It is
immediately obvious that multiple combinations of $f_{eq}$ and $I_{b}/I_{n}$ are consistent with
the observed broad-to-narrow ratios.  This makes the unambiguous, simultaneous determination of 
$f_{H^{0}}$ and $f_{eq}$ from Fig.~5 impossible.  In an earlier calculation, Laming (2000) obtained
similar shapes for the broad-to-narrow ratio curves, the main difference being that our curves
flatten out and then decline slightly at high equilibrations.  The reason for the disagreement is that 
unlike our models, the Laming (2000) calculation maintains a constant shock speed while varying
$f_{eq}$.  The slight downturn in our computed narrow ratio curves is caused by
the decline of the H$^{+}$$-$H$^{0}$ charge exchange rate at high equilibrations (i.e., high
shock speeds).   As a result, the hot neutral density peaks at progressively smaller values
behind the shock as $v_{S}$ increases beyond 3000 \kms.

Fortunately, we can use the He~I/He~II and He~I/H$\alpha$ flux ratios to break the degeneracy in Fig.~5.
Since every He atom in the postshock ionization zone will spend some time as He$^{+}$ (the preshock
He$^{++}$ fraction typically amounts to only a few percent), the predicted He~II $\lambda$4686
flux is {\it insensitive} to the preshock He ionization fraction.  Furthermore, as demonstrated by
Hartigan (1999) and Laming \etal\,(1996), the He~II $\lambda$4686 flux is also insensitive to the
degree of electron-ion temperature equilibration behind the shock (a result that is confirmed
by our models).   Therefore, we can use the He~I/He~II ratio to constrain the preshock He$^{0}$
fraction.  Once this quantity is known, we can then use the He~I/H$\alpha$ ratio to constrain the
preshock H$^{0}$ fraction. This will enable us to determine the appropriate choice of broad-to-narrow ratio from 
Fig.~5.  From our shock models we estimate that regardless of the assumed equilibration, 90\% of the He~II emission 
is produced within a zone $<$1$^{\arcsec}$ 
thick (assuming a total preshock density of 1 cm$^{-3}$ (Winkler \& Long 1997), shock speed 3000 \kms\, and a distance of 2 kpc).
This is smaller than both the pixel scale (0\farcs51 pixel$^{-1}$) and the slit width (2\farcs0).   Therefore, we believe 
that all of the He~II emission layer was contained within the slit.

In most interstellar shocks where He~I emission is observed (mainly radiative shocks),
the $\lambda$5876 line is brighter than the $\lambda$6678 line.  The $\lambda$6678 emission line arises
from the singlet transition $^{1}P\,$(1s2p)\,$-$$^{1}D\,$(1s3d) and is most efficiently excited
from the ground state (i.e., by collisional excitation).  The $\lambda$5876 line, however, arises
from a triplet transition and is most efficiently excited by cascades from high lying energy levels 
(i.e., from recombination).  Therefore, since recombination is negligible in SN 1006 shocks, we expect
the former line to be significantly brighter than the latter (indeed, $\lambda$5876 emission is not 
detected in our optical spectrum).  In the following analysis we limit our calculations to the
$\lambda$6678 line intensities.

To estimate the He~I/H$\alpha$ ratio for SN 1006, we computed the total 
H$\alpha$ and He~I $\lambda$6678 fluxes through the shock front using the combinations of $f_{eq}$ and
$v_{S}$ listed above.
Calculation of the He~I $\lambda$6678 flux in SN 1006 is relatively straightforward because proton
excitation and ionization of He is negligible (Laming \etal\,1996).  In addition, the only significant contribution to
the He~I flux other than $^{1}S\,$(1s$^{2}$)\,$-$$^{1}D\,$(1s3d) excitation are cascades from the
4f level via $^{1}D\,$(1s3d)\,$-$$^{1}F\,$(1s4f).  We have included both direct excitation and cascade 
contributions to 3d in our He~I $\lambda$6678 flux calculation.  We obtained the electron excitation cross sections from
the convergent close coupling (CCC) results of Fursa \& Bray (1995).  At high energies, we approximated the 
cross sections with a Born-like dependence, scaled to match the CCC values at 0.9 keV.  
Our He ionization balance calculation uses the electron ionization rates of Younger (1981) for He$^{0}$ and He$^{+}$.
We also utilized the CCC cross sections of Fursa and Bray (1995) to compute the collisional excitation rate of He~II $\lambda$4686
electrons (proton excitation is negligible, see Laming \etal\,1996).  Given the low density, the order of 
magnitude lower abundance of He relative to H, and the large line width of He~II, we assumed Case A conditions
in the He~II line, i.e., the radiative excitation of He~II $\lambda$4686 by He~II $\lambda$243 (Ly $\gamma$)
absorption is negligible behind the shock.

The model F(6678)/F(4686) ratios are shown in Fig.~6 for a range of preshock 
He$^{0}$ fractions and temperature equilibrations.  The dashed lines indicate error
bars on the observed line ratio after dereddening.  Since the ratio of the He~I excitation rate 
to the He~I ionization rate depends steeply on electron temperature for $T_{e}\,\lesssim\,$10$^{5}$K, 
the He~I/He~II ratio declines sharply at low values of $f_{eq}$ ($\leq\,$0.01).
It is clear from this figure alone that matching the predicted He~I/He~II 
ratio with the observed value requires a preshock He neutral fraction $\geq\,$0.5 and low 
electron-ion equilibration.  

The predicted He~I/H$\alpha$ ratios are displayed in Fig.~7 for a range of preshock H neutral
fractions.  For simplicity, we assumed $f_{He^{0}}\,=\,$1 in these models (varying this quantity
merely moves the curves up or down).  To facilitate comparison with the broad-to-narrow ratios, we used the same set of neutral
fractions and combinations ($f_{eq}$, $v_{S}$) employed by the models shown in Fig.~5.

(1) Case A in both the broad and narrow components, and (2) Case A in the broad component, Case B in the
narrow component.  
Since the inferred Ly $\beta$ and Ly $\gamma$ optical depths behind Balmer-dominated shocks are often
$\lesssim\,$1 in the narrow component line (Chevalier, Kirshner \& Raymond 1980, Ghavamian 1999,
Ghavamian \etal\,2001), the Case A and Case B line ratio curves mark the largest and smallest values of
He~I/H$\alpha$ attainable for each combination of preshock H and He ionization fraction.   

Comparing Figs. 6 and 7, we find that (1) the observed He~I/He~II and He~I/H$\alpha$ ratios are simultaneously
matched for $f_{H^{0}}\,\sim\,$0.1 and $f_{He^{0}}\,\geq\,$0.7 in the preshock gas, and (2) only
the lowest equilibration models ($f_{eq}\,\leq\,$0.05) can match the observations.  Returning to Fig.~5
and choosing the $I_{b}/I_{n}$ curves for $f_{H^{0}}\,=\,$0.1, we find a match between the observed
and modeled H$\alpha$ and H$\beta$ broad-to-narrow ratios for $f_{eq}\,\leq\,$0.1, or $T_{e}/T_{p}\,\leq\,$0.07.  This
corresponds to a shock speed of 2890$\pm$100 \kms\, (including the measurement uncertainty of the
broad component).  This result is in good agreement with the shock speed estimated by Laming \etal\,
(1996) from ultraviolet line analysis of HUT data: 2600$\pm$300 \kms.  It also agrees with the 
low equilibration inferred by Laming \etal\,(1996): $T_{e}/T_{i}\,\leq\,$0.05.   

Given the intrinsic faintness of the He~II emission, it is useful to compare the measured He~II $\lambda$4686
surface brightness with the value inferred from observation of the He~II $\lambda$1640 line by HUT (Raymond \etal\,1995).
Assuming Case A conditions in the He~II lines (reasonable due to the low density of He), the predicted He~II
$\lambda$4686 flux in the RC Spectrograph slit is
\begin{equation}
F(4686)\,=\,\frac{\nu_{4686}}{\nu_{1640}}\,\frac{\gamma_{4686}}{\gamma_{1640}}\,\frac{F_{RCSpec}(H\alpha)}{F_{HUT}(H\alpha)}\,
\,F_{HUT}(1640)
\label{eqn1}
\end{equation}
where $\gamma_{4686}$ is the number of $\lambda$4686 photons per incident He$^{+}$ ion and $\gamma_{1640}$ is
the corresponding quantity for the $\lambda$1640 line.  We have used our H$\alpha$ image (the star-subtracted version of Fig.~1) to scale 
from the He~II $\lambda$4686 flux expected in the HUT aperture to the value expected in the RC Spectograph slit.
This scaling is valid if the H neutral fractions are roughly equal in the two filaments (reasonable given
their close proximity and comparable surface brightnesses).  From the H$\alpha$ image we estimate 
$F_{RCSpec}(H\alpha)/F_{HUT}(H\alpha)\,\approx\,$0.13.  Next, from the calculations of Hartigan (2000) we
estimate $\gamma_{4686}$/$\gamma_{1640}\,\approx\,$0.1 for 10$^{5}\,K\,\leq\,T_{e}\,\leq\,$10$^{7}$\,K.  Inserting
these values into Eqn. \ref{eqn1} and using $F_{HUT}(1640)$ from Raymond \etal\,(1995, who also assumed $E(B\,-\,V)\,=\,$0.11),
we find $F(4686)$\,=\,(1.1$\pm$0.2)$\times$10$^{-15}$ \fl.  This agrees well with the $\lambda$4686 flux we derive from our
data, (9.4$\pm$6.4)$\times$10$^{-16}$ \fl.  

\section{DISCUSSION}

In their detailed imaging study of SN 1006, Winkler \& Long (1997) found that X-ray emission in the NW peaks
approximately 12$^{\arcsec}$ behind the optical filaments.
Using nonequilibrium ionization models convolved with the ROSAT HRI response function, they found that
a total preshock density $\sim\,$1 cm$^{-3}$ was required to match the spatial distribution of the X-ray
emission.    Combined with our estimated H preshock neutral fraction, this implies a preshock neutral
density $n_{H^{0}}\,\sim\,$0.1 cm$^{-3}$ with $\tau_{Ly\,\beta}\,\sim\,$0.5 in the narrow component.
The inferred value of $\tau_{Ly\,\beta}$ will be even lower if Winkler \& Long (1997) overestimated
the total preshock density.  The dereddened narrow Balmer decrement is 3.35$\pm$0.14, significantly lower
than the value of 4.3$-$6 observed
in Tycho's SNR (Ghavamian 2001).  The absorption of narrow component Lyman photons within
the shock structure scales linearly with $f_{H^{0}}$, as long as $f_{H^{0}}$ is
small ($\lesssim\,$0.5).  This enhances the H$\alpha$ 
emission more than H$\beta$, steepening the narrow component Balmer decrement.   Therefore, the modest
decrement we observe in SN 1006 is consistent with our low derived Lyman optical depth.

\subsection{The Preshock Ionization Fractions}

The detection of the He~I $\lambda$6678 line indicates that a significant fraction of the preshock
He is neutral.  The collisional excitation of He~II $\lambda$304 behind the shock
certainly produces some photoionization and heating of the ambient gas (see, for example, Tycho's SNR, 
Ghavamian \etal\,2000).  Since the photoionization rate of He$^{0}$ by $\lambda$304 photons is nearly an order of magnitude 
higher than the photoionization of H$^{0}$ (Ghavamian \etal\,2000), some additional mechanism is required which
can ionize ambient H$^{0}$ more effectively than He$^{0}$.  There are several possibilities which we discuss in
turn below.

\subsubsection{Photoionization From the Reverse Shock?}

One process which may generate high H ionization fractions in the ambient medium while leaving most He neutral
is photoionization by O~II$-$O~VIII emission from the outermost layer of shocked ejecta (e.g., Hamilton \& Fesen
1988, hereafter HF88).  The strongest of these lines have energies lying between the H and He$^{0}$ ionization thresholds.
Using the calculations of HF88 and the He~II $\lambda$304 photoionization models of
Ghavamian \etal\,(2000), we can estimate the ratio of the H photoionization rate to the He
photoionization rate in the interstellar medium due to radiation from the blast wave and shocked ejecta.  
Since each O atom entering the reverse shock will pass
through all ionization stages between O$^{++}$ and O$^{+7}$, the ratio of photoionizations $\alpha_{H/He}(R)$ at
a distance $R$ from the remnant center is 
\begin{equation}
\alpha_{H/He}(R) \, \approx \, 
\frac{\sum_{i}\,\gamma(\lambda_{i})\,\sigma^{H}(\lambda_{i})\,+\,
\eta\,\,\gamma(304)\,\sigma^{H}(304)  }
{\sum_{i}\,\gamma(\lambda_{i})\,\sigma^{He}(\lambda_{i})\, +  \,
\,\eta\,\gamma(304)\,\sigma^{He}(304)}
\label{alpha}
\end{equation}
where 
\begin{eqnarray}
\eta\,=\,\frac{\frac{N_{bw}(He)}{(R\,-\,R_{bw})^{2}}}{\frac{N_{ej}(O)}{(R\,-\,R_{ej}(O))^{2}}} \nonumber
\end{eqnarray}
Here, $\gamma(\lambda_{i})$ is the number of ionizing photons per ion of wavelength $\lambda_{i}$  
(the Bethe parameter as tabulated by HF88 for the reverse shock), $\gamma(304)$
is the Bethe parameter for He~II $\lambda$304 photons from the blast wave ($\approx\,$1.1 for $v_{S}\,>\,$2000 \kms,
Ghavamian \etal\,2000),
$\sigma^{H}$ is the H$^{0}$ photoionization cross section from Osterbrock (1989) and $\sigma^{He}$ is the
corresponding cross section for He$^{0}$ (Reilman \& Manson 1979).  The current O-layer and blast wave 
radii are $R_{ej}(O)$ and $R_{bw}$, respectively, while $N_{ej}(O)$ is the number of shocked O
atoms in the ejecta and $N_{bw}(He)$ is the number of He atoms swept up by the blast wave.  In our calculation we
have assumed that the ionization state of the ISM is negligibly affected by recombination.
HF88 found that the relative masses of Fe, Si and O predicted by carbon deflagration 
model CDTG7 (Woosley \& Weaver 1987) matched the observed Fe~II $\lambda\lambda$2586, 2599 absorption 
profile in SN 1006 best.  Therefore, using the shocked O ejecta mass of 0.41 $M_{\sun}$ and blast wave mass of 4.3$M_{\sun}$
from model CDTG7, $R_{ej}(O)\,=\,$6 pc and $R_{bw}\,=\,$8.3 pc from HF88 and neglecting radiative transfer effects 
on the photoionizing radiation, we
find $\alpha_{H/He}\,\sim\,$0.1 for a parcel of interstellar gas just outside the blast wave radius.  This result suggests that
reverse shock photoionization cannot produce our derived low H neutral fraction.  The 
smallness of $\alpha_{H/He}$ is mainly due to that fact that (1) there are nearly an order of magnitude more He atoms
in the blast wave than there are O atoms in the reverse shock, and (2) since $R_{ej}(O)\,<\,R_{bw}$, a given parcel of interstellar gas
sees a larger emitting area from the blast wave than from the reverse shock.  

A more detailed, time dependent calculation would be desirable.  However, we note that the reverse shock has already propagated
through the entire O-layer, while the blast wave continues to accumulate interstellar
material (and hence He atoms).  Therefore, the ratio $M_{ej}(O)/M_{bw}$ is decreasing as the supernova remnant evolves and the blast wave
is becoming the increasingly dominant photoionization source for the ISM around SN 1006.  Finally, we note that even if we had included
photoionization from the Si-rich and Fe-rich layers in Eqn. \ref{alpha}, our conclusions would remain unchanged since most of the
EUV/X-ray lines from Si and Fe will ionize He much more effectively than H.

\subsubsection{Relic Ionization from the Supernova Explosion?}

An interesting source of photoionization is the EUV flash produced when photons escape from the
detonating white dwarf (Type Ia SN).  Using the EUV burst models presented by Blinnikov \& Sorokina (2000) and Sorokina \& Blinnikov (2001,
in preparation), we estimate $\sim\,$10$^{55}-$10$^{56}$ ionizing photons (E$\,>\,$13.6 eV) are released over the integrated
Type Ia SN light curve.  The precise number can vary
by a factor $\sim$5, depending on which detonation model is used in the light curve calculation (see Blinnikov \& Sorokina (2000) 
for model descriptions).  Due to the steepness of the EUV spectrum, the H$^{0}$ ionization rate is much larger than the He$^{0}$ 
ionization rate, a desirable feature.  However, assuming an average ISM density of 0.06$-$1.0 cm$^{-3}$ around SN 1006 and distance of 2 kpc, 
there were $\gtrsim\,$5$\times$10$^{57}$ H atoms contained within the current blast wave radius at the time of the SN explosion.
Even if these H atoms were already partially ionized beforehand, the total ionizing flux from a Type Ia explosion still falls
an order of magnitude short of the value needed to effectively ionize the gas at the current blast wave radius.
The disparity is worsened if we include radiative transfer 
loss of the released photons as they (1) propagate through the ISM, and (2) possibly encounter a circumstellar envelope
left by a nondegenerate donor star prior to the explosion (Cumming \etal\,1996).  Therefore, regardless of whether the EUV emission 
ionizes H more effectively
than He, there are not enough photons produced in the explosion to explain our inferred neutral fractions.  This result
also agrees with the SN flash ionization calculations of Cheng \& Bruhweiler (1996).

\subsubsection{Photoionization from Galactic Background Radiation?}

If neither the supernova event nor the postshock emission from the supernova remnant can account for our
deduced H$^{0}$ and He$^{0}$ fractions, we must consider nonlocal sources of ionizing radiation.  Given the lack
of late O and early B stars in the vicinity of SN 1006, the most promising source is the Galactic ionizing
background.  Reynolds (1984) observed the H$\alpha$ recombination line toward a sample of high Galactic 
latitude pulsars ($\mid b\mid\,>\,$5) at distances of 2$-$3 kpc.  He found that Lyman continuum radiation from O stars
in the Galactic plane was the most likely source of ionization 
in the diffuse ISM.  Since the Lyman continuum is produced by recombination, it declines nearly
exponentially beyond each ionization edge (Brussard \& Van De Hulst 1962).  Therefore, incident Lyman continuum
photons will ionize H far more effectively than He, a desirable property.

The remaining question is whether the Galactic Lyman continuum flux is high
enough at the position of SN 1006 ($b\,=\,$+14.6$^{\circ}$) to produce the necessary amount of H ionization. 
Assuming a density of 1 cm$^{-3}$ around SN 1006, the recombination rate
for 5000~K$\lesssim\,$T$\lesssim$10,000~K is $\sim$5$\times$10$^{-13}$ s$^{-1}$ (Osterbrock 1989).  Therefore, obtaining an H neutral 
fraction $\sim$0.1 requires a photoionization rate $\sim$5$\times$10$^{-12}$ s$^{-1}$.  
Dove, Shull \& Ferrara (2000) estimated a Lyman continuum flux $\sim\,$5$\times$10$^{7}$ photons cm$^{-2}$ s$^{-1}$ from O stars in the
Galactic disk.  In their time dependent radiative transfer calculation for the ionizing continuum, they estimated
that roughly 3\%$-$6\% of the Lyman continuum photons escape through superbubbles and Galactic chimneys into the halo of the
Milky Way, producing the ``Reynolds Layer'' of diffuse H$\alpha$ emission (Reynolds 1984).  If
the Lyman photons propagating into the halo suffer no further attenuation, the corresponding H ionization rate 
in the vicinity of SN 1006 is approximately
$F_{Ly}\,\sigma^{H}$(912) $\sim$1$-$2$\times$10$^{-11}$ photons s$^{-1}$, more than enough to produce the necessary
H ionization around SN 1006.  

If the Galactic Lyman continuum background pre-ionizes most of the ambient H on the NW side of SN 1006,
the question arises of why the ambient H on the eastern half of Tycho's SNR (also the remnant of a Type Ia explosion)
was found to be mostly neutral (Ghavamian \etal\,2000).  First, the eastern half of Tycho's SNR is likely interacting
with the outer edge of a warm H~I cloud (Reynoso \etal\,1999, Ghavamian \etal\,2000).  Second, the presence of large structures in
the Galactic disk (H~I clouds, superbubbles, molecular cloud complexes, etc.) produces large variations in the Lyman continuum flux
from position to position and time to time (Dove, Shull \& Ferrara 2000).  Since Tycho's SNR is close to the Galactic 
plane ($b\,=\,$+1.4$^{\circ}$), it is possible that this remnant is shielded from the Lyman continuum
emission by the many dense clouds in the Galactic disk.  SN 1006, on the other hand, lies nearly 500 pc above the
Galactic plane.  Therefore, neutral H near SN 1006 may ``see'' a larger time averaged Lyman continuum
than the neutral H around Tycho's SNR.  

\subsection{Comparison With Ionization Estimates of the Local Bubble}

An interesting comparison can be drawn between our derived H and He neutral fractions and values estimated
for the local ($\lesssim\,$200 pc) ISM.  For example, using EUVE spectra of 29 DA white dwarfs, Wolff \etal\,(1999) 
estimated $f_{He^{+}}\,\sim$\,0.4 in the Local Bubble, with no evidence of a gradient or other general pattern.
They interpreted this result as evidence for a lack of He-ionizing sources in any particular location.
On the other hand, Wolff \etal\,(1999) found a positive gradient in the H ionization fraction toward the Canis
Major cavity, implicating the B-stars Adara and Mirzam as the photoionizing sources.  In some respects the ionization properties
of the Local Bubble and the ISM near SN 1006 are similar: the H ionization is strongly influenced by EUV radiation,
while He remains largely unaffected.  The main difference between the two cases, of course, is the lack of a local
H-ionizing source near SN 1006.  Our derived He ionization fraction ($\leq\,$0.3) for SN 1006 is lower than
the average value of 0.4 measured by Wolff \etal\, (1999), but there are individual cases from their WD sample
with He ionization fractions similar to ours.  

\subsection{Comparison With Earlier Results}

Our optical analysis lends strong support to the picture developed by earlier X-ray, UV and optical
studies of SN 1006.  Multi-epoch optical imagery of SN 1006 has yielded a proper motion of
0.3$^{\arcsec}$$\pm$0.04$^{\arcsec}$ yr$^{-1}$ (Long, Blair \& van den Bergh 1988) for the Balmer filaments
in the NW.  Combined with our
estimated shock speed this gives a distance of 2.0$^{+0.4}_{-0.3}$ kpc, agreeing well with the 
value 1.8$\pm$0.3 obtained by Laming \etal\,(1996).  The SN 1006 blast wave is the fastest
shock we have thus far modeled with our Balmer-dominated shock code.  In modeling Balmer spectra from
the Cygnus Loop ($v_{S}\,\sim\,$300 \kms), RCW 86 ($v_{S}\,\sim\,$600 \kms) and Tycho's SNR ($v_{S}\,\sim$\,2000
\kms), we noted that the equilibration $f_{eq}$ declined from near 1 for the slowest shocks to 
$\lesssim\,$0.2 for the fastest shocks (Ghavamian 1999, Ghavamian \etal\,2001).  The shocks in 
SN 1006 are closer to 3000 \kms\, and are matched by an even lower $f_{eq}$, further evidence of an inverse
correlation between shock speed and equilibration.  The case for low equilibration at high Mach numbers has
recently received further support from X-ray observations of SN 1987A.  In their analysis of high resolution 
HETG data from {\it Chandra}, Michael \etal\,(2001) found that a high blast wave speed (3500 \kms)
along with a low electron temperature ($f_{eq}\,<\,$0.1) were required to match the X-ray profiles
of the SN 1987A spectrum.

\subsection{Implications for Collisionless Shock Physics}

There are few numerical simulations available of high Mach number, collisionless shocks.  Cargill \& 
Papadapoulos (1988) modeled the electron-ion heating in a shock moving perpendicular to the ambient magnetic
field, with $M_{A}\,=\,$50 ($M_{A}\,\equiv\,v_{S}/(c_{s}^{2} + v_{A}^{2})^{1/2}$ is the Alfv\'{e}nic Mach number, 
where $c_{s}$ is the sound speed and $v_{A}$ is the Alfv\'{e}n speed of the preshock gas).  In their hybrid simulation the electrons were
treated as a resistive fluid, while the ions were treated as individual particles. Cargill \& Papadapoulos (1988)
found that two-stream plasma instabilities
produced by ions reflected upstream resulted in $T_{e}/T_{i}\,\sim\,$0.25 at the shock front.  Shimada \& Hoshino
(2000) tried an approach where both electrons and ions were treated as particles.  At the highest
Mach numbers covered by their simulations, they predicted that the equilibration would {\it increase}
with Mach number.  In their fastest simulated shock ($M_{A}$\,=\,20), they found $T_{e}/T_{i}\,\sim$\,0.2.
Their curve of $T_{e}/T_{i}$ vs. $M_{A}$ flattens considerably as $M_{A}\,\rightarrow\,$20; extrapolating
this curve to $M_{A}\,\gtrsim\,$200 leads to significantly larger
equilibration than implied by the ultraviolet (Laming \etal\,1996) and 
optical (this paper) observations of SN 1006. 

During the last few years, a considerable effort has been made in modeling collisionless heating
processes in solar wind shocks ($M_{A}\,\leq\,$20).  Hull \& Scudder (2000) combined an electron fluid treatment with 
constraints from ISEE 1 data of Earth's bow shock to model the electron-ion equilibration process.
For the first time, they were able to match the observed electron temperature jumps in fast shocks propagating
both parallel and perpendicular to the magnetic field.  These models emphasize the role of coherent (DC)
electric forces in the collisionless heating, rather than wave-particle interactions.  The advantage of
the Hull \& Scudder (2000) treatment is that it draws upon in situ observations of real collisionless
shocks rather than computer simulations.  Both these models and the observations (Schwartz \etal\,1988)
predict $T_{e}/T_{i}\,\sim\,$0.25 for shocks in the solar wind.   Although we are unable to perform similar detailed
observations of supernova shocks, extending the models of Hull \& Scudder (2000) to much higher Mach numbers
may be an interesting direction for future work on collisionless shocks.

\section{SUMMARY}

The optical spectrum presented here is the deepest yet obtained of the nonradiative filament
along the northwest rim of SN 1006.  Lines of H$\alpha$, H$\beta$, H$\gamma$, H$\delta$, He~I $\lambda$6678
and possibly He~II $\lambda$4686 have been detected$-$more lines than in the spectrum of any other
purely nonradiative SNR filament.  We have detected both broad and narrow component emission lines of
H$\alpha$, H$\beta$ and H$\gamma$.  The broad and narrow components result, respectively, from
fast neutrals produced by charge exchange and slow neutrals that are collisionally excited.

The broad component widths, the ratios of broad to narrow flux in H$\alpha$ and H$\beta$, and
the Balmer line strengths relative to He~I and He~II enable us to determine the important
parameters for the shock.  We find a shock speed of 2890$\pm$100 \kms\,and a very low
degree of electron-ion equilibration, $T_{e}/T_{p}\,\leq\,$0.07, both consistent
with earlier measurements.  In addition, our models can match the observations only if
the preshock H is mostly ionized ($\sim\,$90\%) and preshock He is mostly neutral ($\gtrsim\,$70\%).  In seeking
an explanation for this ionization structure, we have concluded that neither radiation from the blast wave
nor radiation from the reverse shock can effectively ionize the ambient H surrounding SN 1006.
We also find that the EUV flash from the supernova explosion is incapable of producing the
necessary H ionization.  The most plausible explanation for the high H ionization fraction
seems to be photoionization from the Galactic Lyman continuum.  Finally, comparing our derived
electron-ion equilibration with values derived for other nonradiative SNRs reinforces the
earlier conclusion that the fastest shocks have the least equilibration.  A thorough understanding
of this trend remains as a challenge for the study of collisionless shock physics.

The authors gratefully acknowledge the support staff and general hospitality of CTIO, where
various parts of this work were completed.  We would like to thank P. Lundqvist for helpful
correspondence on the EUV bursts from SNe and E. I. Sorokina for providing us with EUV
fluxes from her emission calculations of Type Ia SNe in advance of publication.
P. G. would like to thank J. P. Hughes for valuable discussions, and Marc Hemsendorf for valuable
advice on optimizing the codes used in this paper.  The authors wish to acknowledge their sources
of financial support: P. G., through Chandra grants GO0-1035X and GO1-2052X to Rutgers University;
P. F. W., through NSF grant AST-9618465 and NASA grant NAG 5-8020; J. C. R., through FUSE grant
NAG5-10352; and K. S. L., through Chandra grants GO0-1120X and GO1-2058A.

\clearpage

\begin{figure}
\plotone{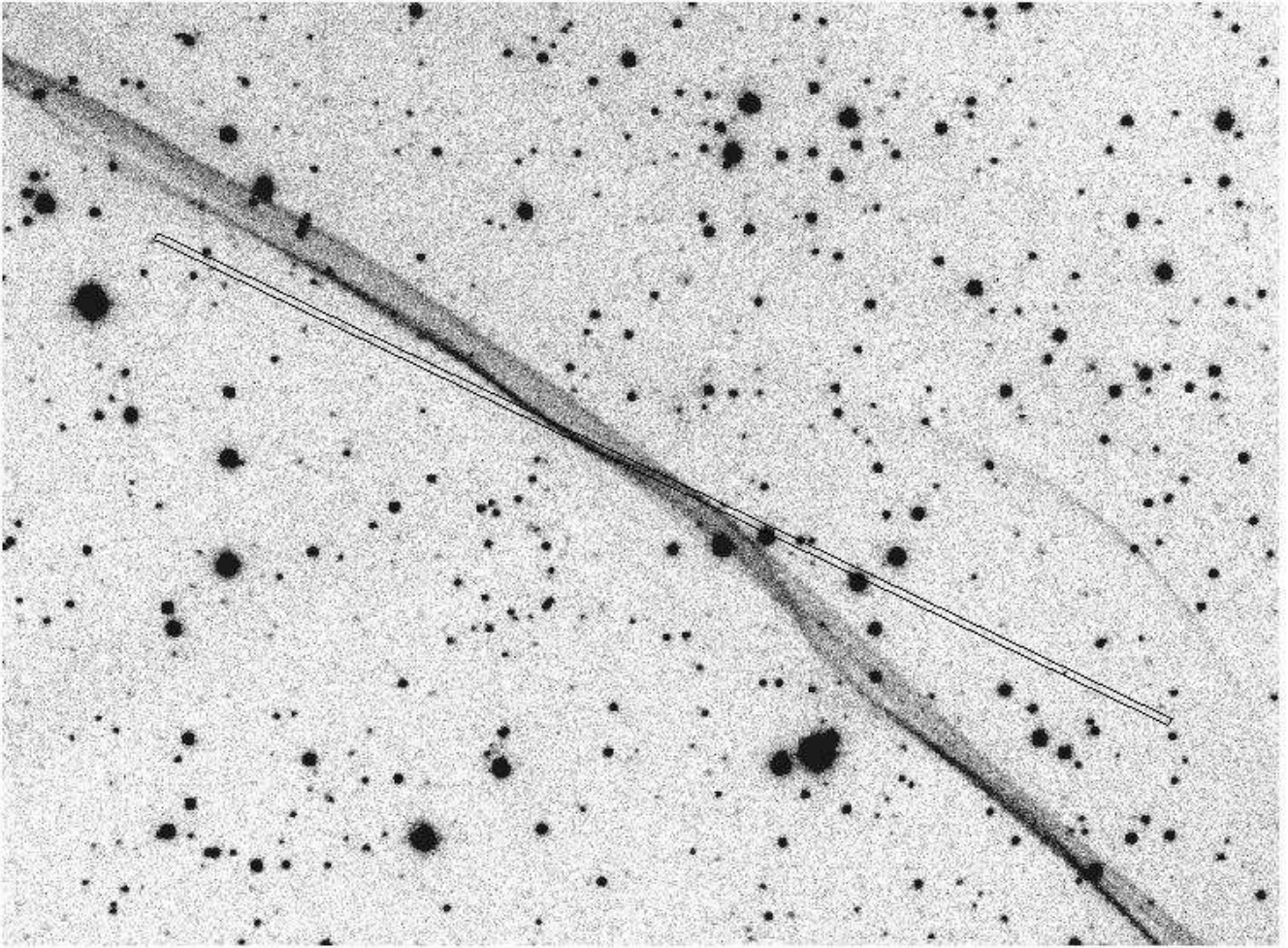}
\figcaption{A narrow band H$\alpha$ image of NW SN 1006, acquired from the CTIO 0.9-m telescope in
1998 June, shortly before the spectroscopy reported here.   The location and PA of the 
2$^{\arcsec}\,\times\,$5$^{\arcmin}$ RC Spectrograph slit is marked.  North is at the top, East is to the left. }
\end{figure}

\begin{figure}
\plotone{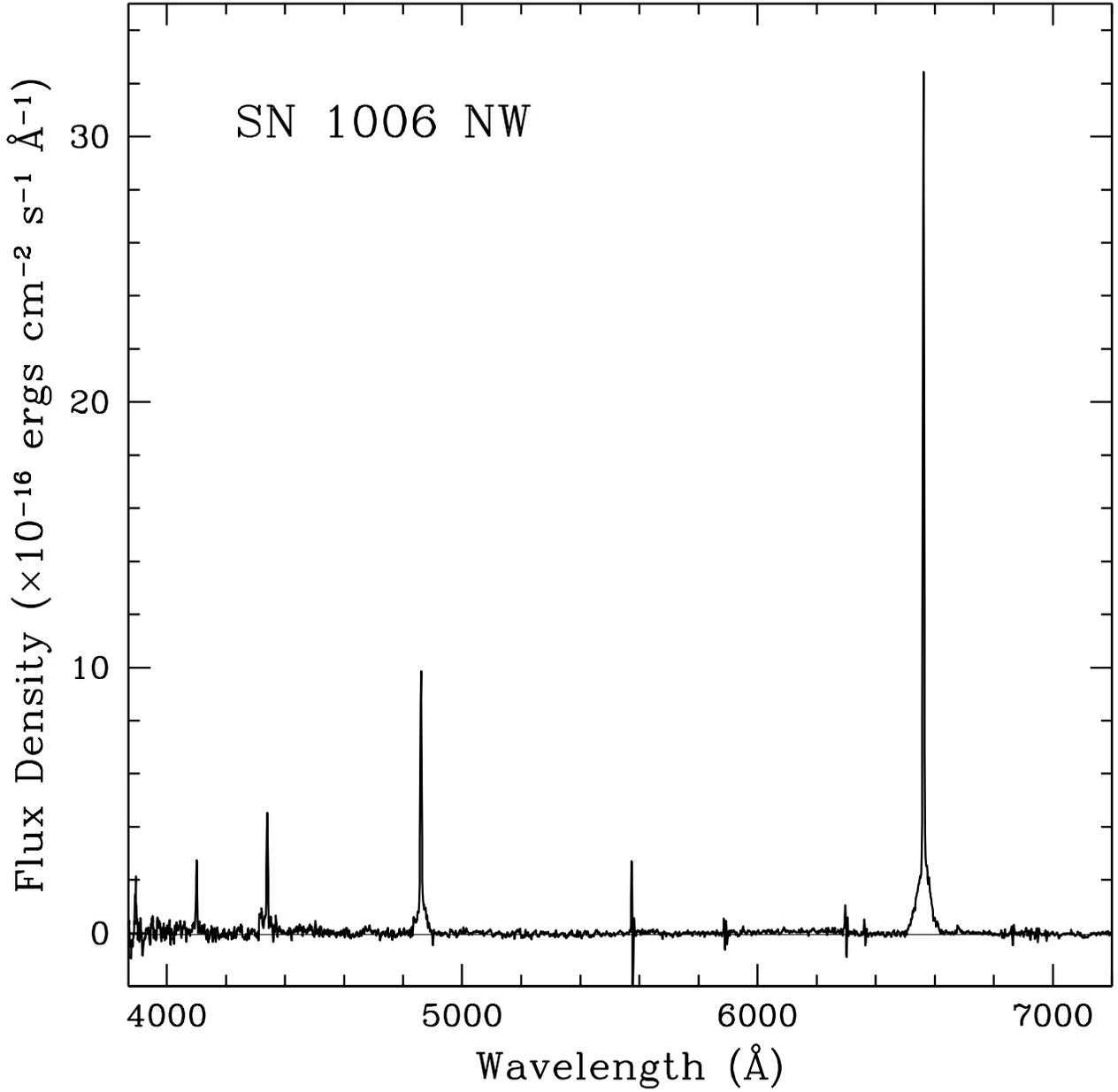}
\figcaption{Full one-dimensional spectrum of the NW SN 1006 Balmer filament.  Narrow emission lines
of H$\alpha$, H$\beta$, H$\gamma$ and H$\delta$ are detected, along with broad emission lines
in H$\alpha$, H$\beta$, and possibly H$\gamma$. The sharp feature near the middle
of the spectrum is an artifact of the sky subtraction.}
\end{figure}

\begin{figure}
\plotone{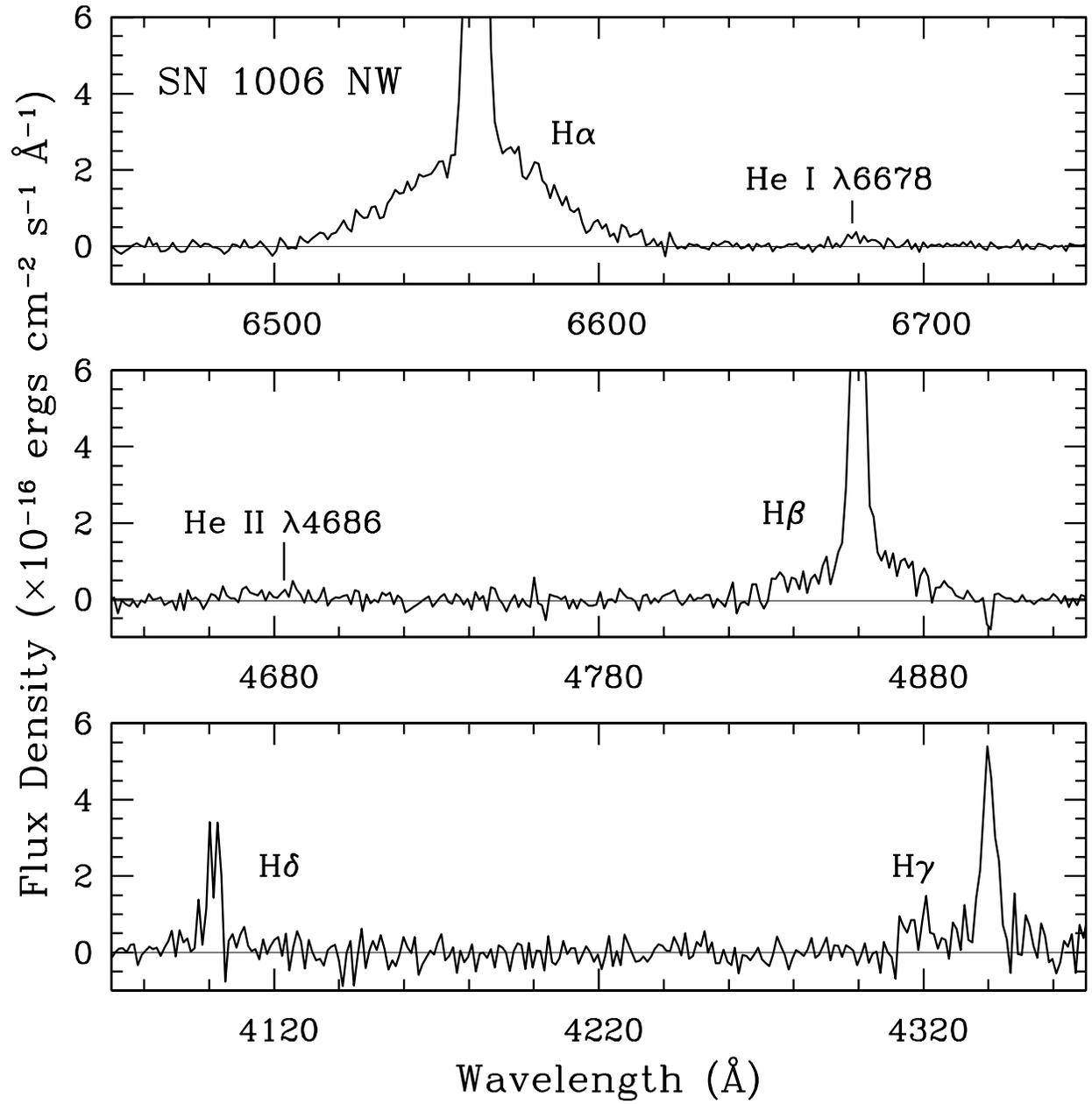}
\figcaption{Closeup view of the NW SN 1006 Balmer spectrum.  No smoothing has been applied.
Among the newly detected lines is He~I $\lambda$6678.  
There may also be a weak detection of the He~II $\lambda$4686 line.}
\end{figure}

\begin{figure}
\plotone{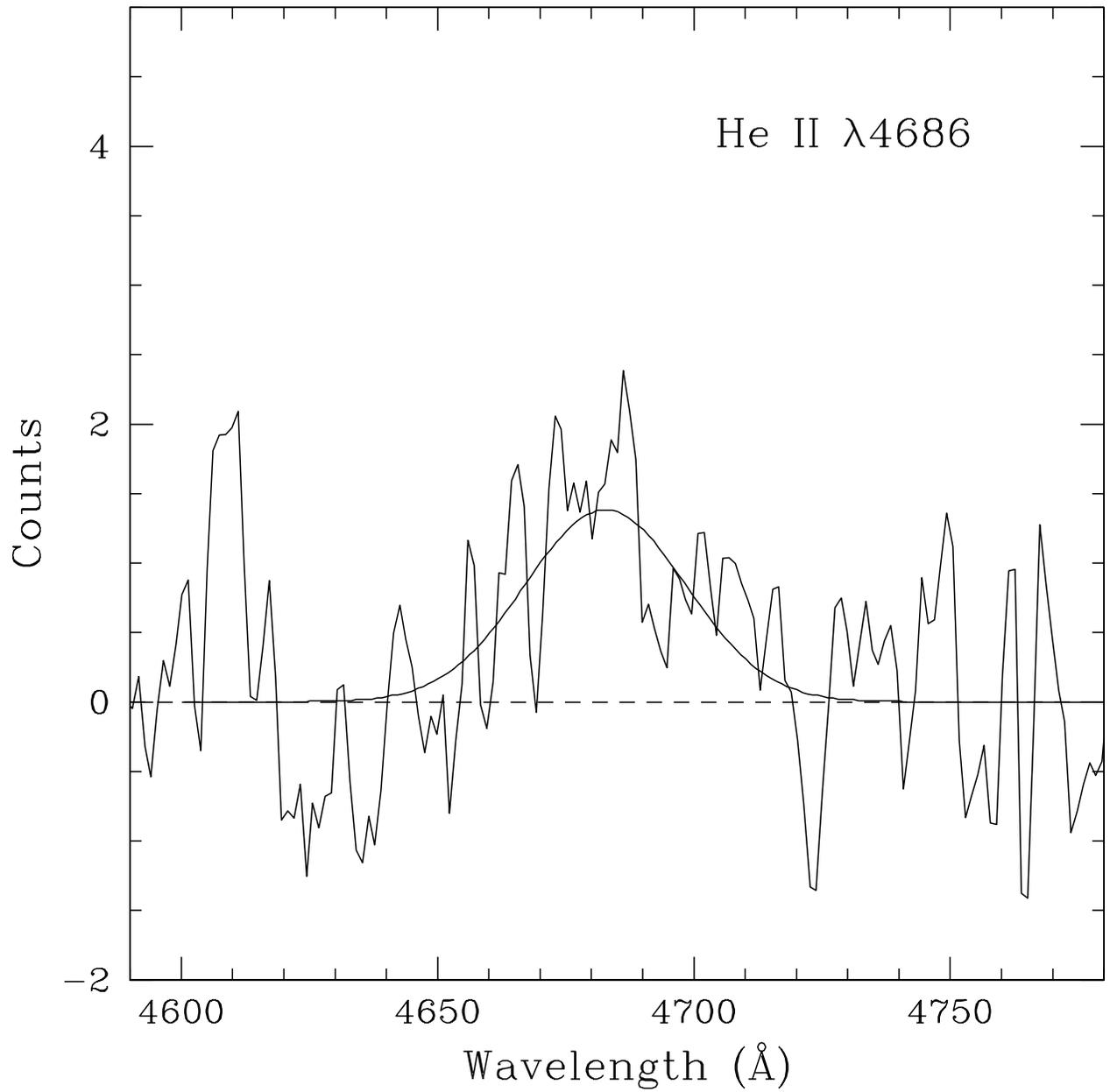}
\figcaption{The He~II $\lambda$4686 line profile.  The fit has been performed assuming the He~II line
width equals that of H$\alpha$. }
\end{figure}

\begin{figure}
\plotone{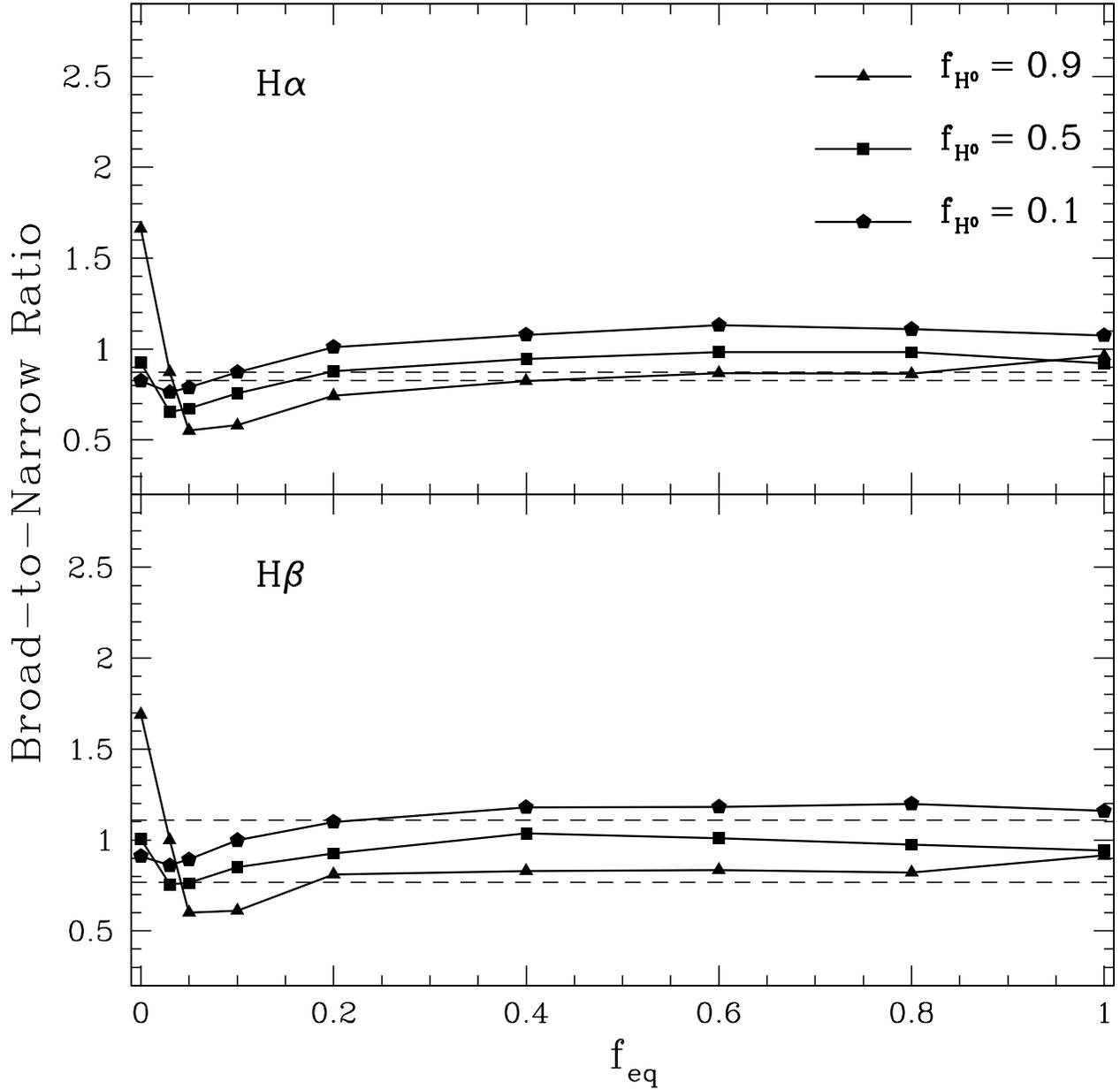}
\figcaption{The calculated H$\alpha$ and H$\beta$ broad-to-narrow ratios vs. the electron-ion
equilibration fraction $f_{eq}$ at the shock front.  The dashed horizontal lines mark the upper and lower limits of the observed 
broad-to-narrow ratios.  Curves of constant preshock neutral fraction are shown for cases where the
preshock H is mostly neutral, half neutral and mostly ionized. }
\end{figure}

\begin{figure}
\plotone{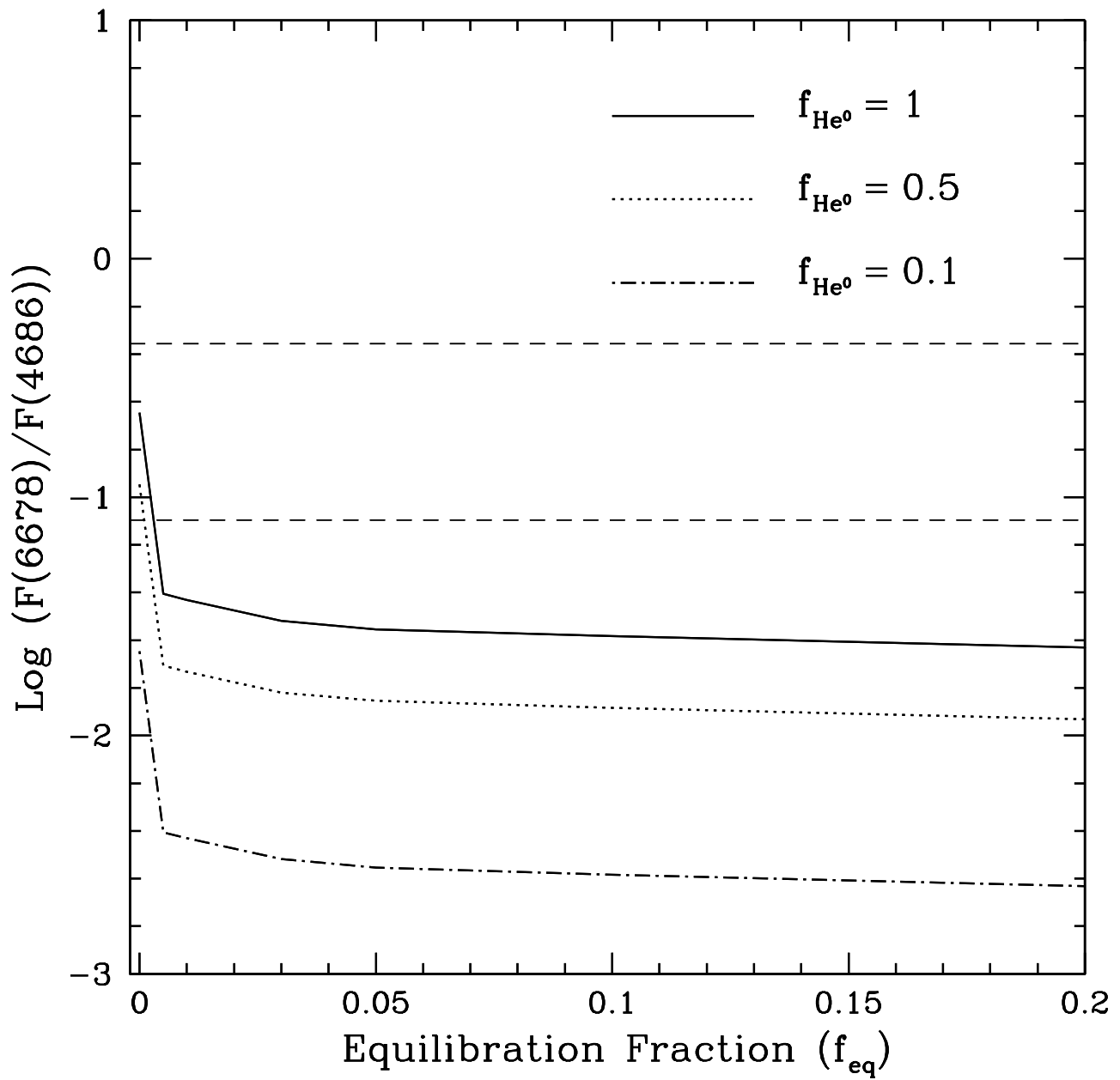}
\figcaption{The observed vs. predicted He~I/He~II line ratio, shown for a range of preshock
He neutral fractions. The observed range of the He~I/He~II ratio is indicated by the
horizontal dashed lines.  The model ratios have been computed assuming that neutral H in the narrow
component is optically thin to Ly $\beta$ photons behind the shock (Case A). }
\end{figure}

\begin{figure}
\plotone{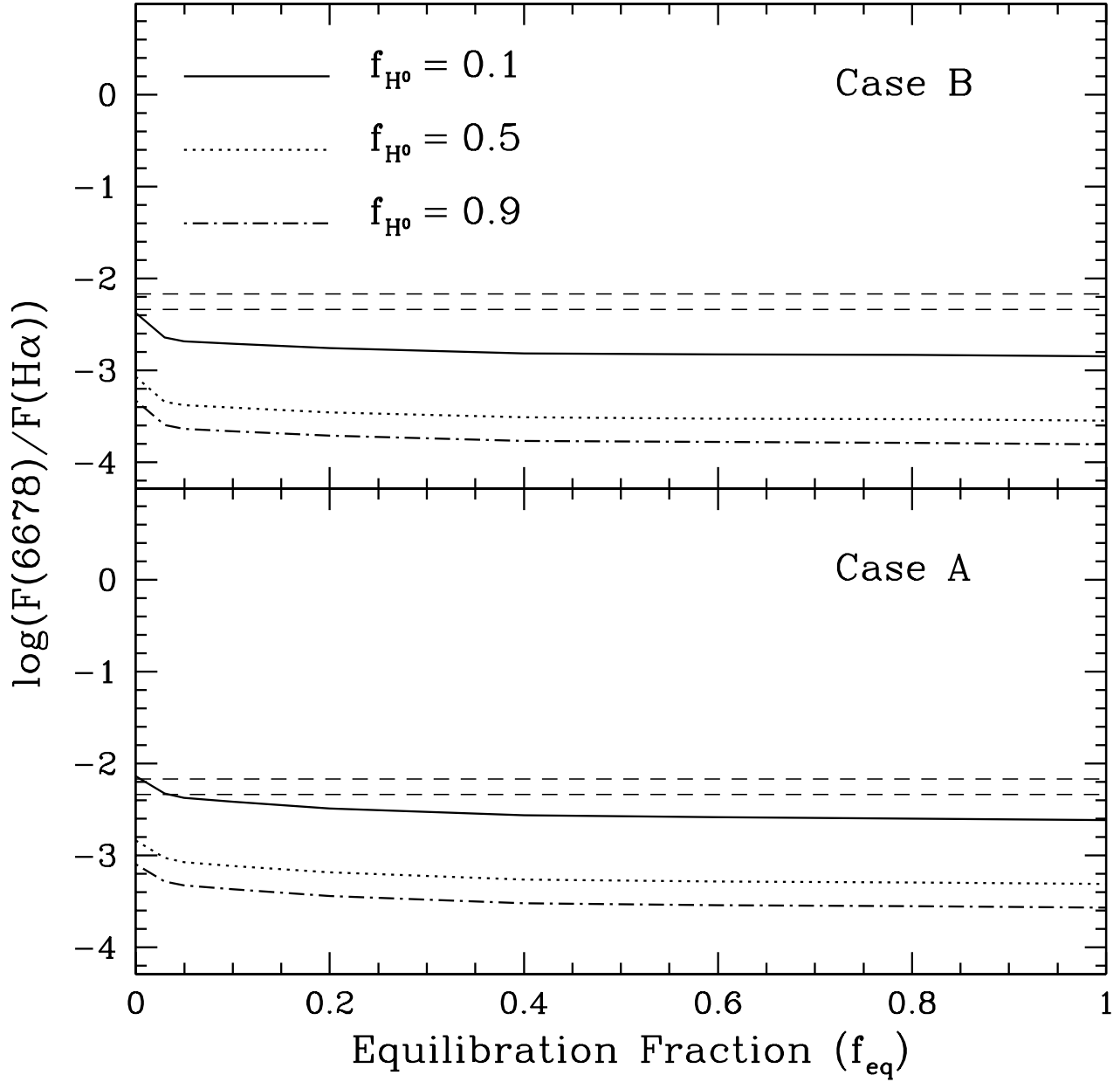}
\figcaption{The dependence of the He~I/H$\alpha$ ratio on $f_{eq}$, shown
for a sampling of preshock H neutral fractions.  The preshock He is assumed to be fully neutral.
The dashed lines indicate the observed range of the He~I/H$\alpha$ flux ratio.
The ratios in the upper figure are computed assuming that neutral H
in the narrow component is optically thick to Ly $\beta$ behind the shock (Case B), while the
ratios in the lower figure assume that neutral H in narrow component is optically thin to Ly $\beta$ behind the shock (Case A). }
\end{figure}

\begin{deluxetable}{cccccc}
\tablenum{1}
\tablecaption{Observed Surface Brightnesses For NW SN 1006\tablenotemark{a}}
\tablehead{
\colhead{Line} &     
\colhead{F($\lambda$)$_{obs}$} & \colhead{I($\lambda$)\tablenotemark{b}} & \colhead{Broad to Narrow Ratio} &
\colhead{Velocity Width (\kms) }   }
\startdata
H$\alpha$ & 355.3$\pm$34.8 & 318.1$\pm$31.2 & 0.84$^{\,+0.03}_{-0.01}$  &   2290\,$\pm$\,80  \\
    &  &  &  &  \\
H$\beta$ & 100\tablenotemark{c} & 100\tablenotemark{d} &  0.93$^{\,+0.18}_{-0.16}$  &  2290 (fixed)\\
    &  &  &  &  \\
H$\gamma$  & 36.4$\pm$7.4 & 38.1$\pm$7.7  &  0.63$^{+0.53}_{-0.25}$   &   2290 (fixed)\\
H$\delta$  & $>$\,16.8 & $>$\,18.3  &  \nodata    &   215 (fixed)\tablenotemark{e}\\
He~I $\lambda$6678 & 2.0$\pm$0.4  & 1.9$\pm$0.44  &  $-$\tablenotemark{f}   &  215 (fixed)\\ 
He~II $\lambda$4686 & 6.6$\pm$4.5  & 7.0$\pm$4.8  &   $-$   &  2290 (fixed)\\ 
\tablenotetext{a}{Balmer line surface brightnesses are quoted for the combined broad and narrow component emission from
a 2$^{\arcsec}\times$51$^{\arcsec}$ section of filament. The error bars are 1$\sigma$.}
\tablenotetext{b}{Dereddened intensities assume E(B$-$V)\,=\,0.11$\pm$0.02 (Schweizer \& Middleditch 1980). }  
\tablenotetext{c}{Observed total H$\beta$ surface brightness of 9.0$\times$10$^{-17}$ \sb.}
\tablenotetext{d}{Dereddened total H$\beta$ surface brightness of 1.3$\times$10$^{-16}$ \sb.}
\tablenotetext{e}{Narrow component line widths are fixed at 205 \kms (the instrumental resolution).}
\tablenotetext{f}{The He~I line is intrinsically narrow, while the He~II line width is intrinsically broad.}
\enddata
\end{deluxetable}

\end{document}